\documentstyle[12pt,epsfig]{article}
\author{ M.A. Per
\footnote{maper@jazzfree.com}
\ A.J. Segu\'{\i}
\footnote{segui@posta.unizar.es}\\
Departamento de F\'{\i}sica Te\'orica. Facultad de Ciencias.\\
Universidad de Zaragoza. 50009-Zaragoza, Spain}
\title{Encoding the scaling of the cosmological variables with the Euler Beta function}
\begin{document}
\maketitle
\begin{abstract}
We study the scaling exponents for the expanding isotropic flat cosmological models.
The dimension of space, the equation of state of the cosmic fluid and
the scaling exponent for a physical variable are related by the Euler Beta
function that controls the singular behavior of the global integrals.
We encounter dual cosmological scenarios using the
properties of the Beta function. When we study the integral of the density of entropy
we reproduce the Fischler-Susskind holographic bound.
\end{abstract} 
\vspace{2cm}

\vspace{.5cm}


\newpage

\newcommand{\be}{\begin{equation}}
\newcommand{\ee}{\end{equation}}
\newcommand{\ci}{\cite}
\newcommand{\la}{\label}

In the present situation of the universe undergoing an expansive,
maybe accelerated \ci{Pelmuter} phase and
characterized by its flat nature as well as its homogeneity and isotropy,
it seems interesting to relate what we observe with what in fact exists today:
We want to compare the integral of the density of physical magnitudes over
our past light cone (what we \emph{see}) with the same densities but now integrated over
the spatial section connected today causally (what \emph{exists} today).
In general an infinite number of foliations are possible
for a given manifold and usually
the restriction to one or other of them is in accordance to physical
properties (mainly symmetries). Two important symmetries are at the core of the description
of the observed universe: covariance and isotropy. The covariant symmetry is incorporated
with a null foliation and it is over these light sheets that can be formulated a covariant
version \ci{bousso} of the holographic principle \ci{'t hooft}.
The isotropy of the spatial sections is closely
related with the horizon problem and inflation \ci{linde}
or some string/M theory property \ci{banks}.

Therefore, let us consider two foliations of a
cosmological manifold of dimension $D=n+1$,
the manifold being a flat expanding Friedman-Robertson-Walker (FRW) universe,
with a power law behavior for the
evolution of the scale factor with cosmic time $R(t) \propto t^{1-1/\alpha}$
with $\alpha$ constant.
One of the two foliations is covariant and is constructed by the superposition of
the past light cones from the big bang;
we take the past light cones for a fiducial observer that is placed at the origin
and that moves with the cosmic fluid and, because our universe satisfies the cosmological
principle, our observer is a generic one.
Each light sheet can be parametrized by a time like coordinate that we
take the proper time of the observer and that coincides with the cosmic time.
The other foliation uses space like $n$-dimensional
surfaces parametrized with the same cosmic time, also referring to the same observer, and
constructed according to the cosmological principle so that are homogeneous and
isotropic surfaces. In this second case we bound, the
otherwise infinite surfaces, with a cosmological horizon.
Here we encounter a difference depending on the type of the expansion; if the expansion
is decelerated, the spatial sections are bounded
by a particle horizon; in the case of an accelerated evolution
it is the event horizon that bounds the spatial regions.
If we restrict to an ideal cosmic fluid with energy density $\rho$
and pressure $p$, then $(2-n)\rho < np$ gives a decelerated evolution
while $(2-n)\rho > np$ gives an accelerated one.
If the equation of state is of the form $p=\omega \rho$
it is the value of $\omega$ which controls the character
accelerated or decelerated of the expansion for the ranges
$-1 \leq \omega \leq (2-n)/n$ and
$(2-n)/n <\omega \leq 1$ respectively. The scaling exponent for the scale factor
is then given by $1-1/\alpha= 2 /n(1+\omega)$.
For the physical or geometric densities that we investigate, a power law evolution with
cosmic time is assumed $q(t)=P t^{m}$ where $P$ gives the dimension and the scale
of the density and will be irrelevant in the present study.
In this situation, our description is formulated with
the following three parameters $n$, $\alpha$ and $m$ and we will be able to restrict them
by physical/geometrical arguments.
In general we encounter that the quotient of the integral of the physical densities
for the two foliations is given by the Euler Beta function and its properties
are used to limit the parameter space for different physical densities.

The general result can be applied to different magnitudes, in particular to the entropy
of the universe in order to implement the holographic principle.
The holographic principle can be considered as a principle underlying the quantum fluctuation of the
geometry and then a window over the quantum gravity theory;  
the universe as a whole must be consistent with this principle and is this aspect that is studied in this article.
When we analyze the entropy for the two foliation we must
be sure on the finite value of the entropy density integral over the light sheets.
This is a necessary condition because otherwise there is no room on the finite horizons
to project the degrees of freedom and consequently the holographic bound
is violated. The horizons represents our ignorance on the events of the manifold and can be related with
the gravitational entropy; so, they are appropriate to encode holographically the information of the bulk.
We are able to restrict the parameters to avoid this infinity.
If the expansion is adiabatic the parameter space is more restricted and we reproduce the
Fischler-Susskind bound \ci{fischler}.

The metric for the flat FRW universe of $D=n+1$ dimensions is given by
\be\la{1}
ds^{2}=-dt^{2}+R^{2}(t) \Big(dr^{2} +r^{2}d\Omega_{n-1}^{2} \Big),
\ee
where $d\Omega_{n-1}^{2}$ is the line element for the $n-1$ dimensional unit sphere
${\mathbf{S}}^{n-1}$.
To measure distances in the previous geometry, because the isotropy condition, we can fix the
direction by fixing the polar angles without loss of generality. We
coordinate the events using proper distances $D(t)=R(t)r$ from the points of the manifold,
with coordinates $(r,t)$, to the origin at the same time $(r=0,t)$ where we place
our fiducial observer.
The proper distance to the origin $r=0$ as a function of cosmic time $t$,
for the null geodesic that at time
$t_{0}$ ($t_{0}>t$) is seen by the observer at $r=0$ is obtained integrating (\ref{1}) with $ds^{2}=0$
with the result ($R(t) \propto t^{1-1/\alpha}$)
\be\la{2}
D(t,t_{0})=\alpha t_{0}   \Big[ \Big( {t \over t_{0} } \Big)^{1-1/\alpha}-
				  \Big( {t \over t_{0} } \Big)  \Big].
\ee
See \cite{nos}. 
Given a density $q(t)$ for a physical magnitude that, due to the
cosmological principle depends only on cosmic time $t$, we can determine its integral
over each of the sections of the null foliation of the universe;
we integrate $q(t)=P t^{m}$ over the past light cone of the event $(D=0, t_{0})$
and we do this from the big bang $t=0$,
\be\la{3}
Q_{null}(t_{0})= \int_{0}^{t_{0}} A_{n-1} D(t,t_{0})^{n-1} P t^{m} dt,
\ee
where $A_{n-1}=2 \pi^{n/2} / \Gamma(n/2)$ is the area of the $n-1$
dimensional unit sphere.
The same density can be integrated over the homogeneous spatial sections
at different cosmic times. If the expansion is decelerated, $\ddot R<0$, and consequently
$0<1-1/\alpha<1$; then the causally connected region at time $t$ is bounded by the
particle horizon whose size is the proper distance to the origin of the
outgoing photon that at the big bang ($t=0$) was at $r=0$; to obtain its
value we must consider $-D(t,t_{0})$ in (\ref{2}) (the minus sign is due to the fact
that we are looking for outgoing photons) and take its limit for $t_{0}$ going to zero ($t_{0}<t$),
\be\la{4}
D_{PH}(t)=\lim_{t_{0} \rightarrow 0} -D(t,t_{0}) = \alpha t.
\ee
If the expansion is accelerated it is the event horizon the border of the spatial
region. In this case $1-1/\alpha>1$ and  we see that the slope of the null ray given
by (\ref2) is a decreasing function of $t$ with maxima at $t=0$;
the straight line tangent to the light cone at the big bang is then a limit
for the photons arriving to $r=0$, it is then the event horizon of the origin.
It is a straight line because we are assuming
that $\alpha$ is constant in the cosmic evolution. The event horizon of the origin
separates light rays with opposite concavity. Taking the limit of
the derivative with respect of $t$ of $D(t,t_{0})$ when $t_{0} \rightarrow 0$
we obtain
\be\la{5}
D_{EH}(t)=\lim_{t_{0} \rightarrow 0}
\Big( { dD(t,t_{0}) \over dt} \Big) t = - \alpha t.
\ee
By the relations (\ref{4}) and (\ref{5}) we see the dual nature of both horizons; when
there is particle horizon there is not event horizon (it is negative), and viceversa. The
absolute value of both of them is the same.
The density of a given magnitude is constant on the spatial sections because they are
constructed in accordance with the homogeneity hypothesis;
then it is trivial to obtain the integral on the section for the physical
magnitude; it will be the product of the density and the volume of
the region bounded by the corresponding horizon at a given time $t_{0}$ (the same
that parameterize the null foliation). Consequently, we have
\be\la{6}
Q_{hor}=V_{n}(\alpha t_{0})^{n} P t_{0}^{m},
\ee
that corresponds for $\alpha>0$ to a decelerated evolution
\footnote{ Strictly $\alpha$ must be bigger than $1$ because we are dealing
with expanding models.}
and the same expression changing
$\alpha$ by $-\alpha$ for an accelerated one. In (\ref{6}) $V_{n}=A_{n-1}/n$ is the
volume enclosed by the $n-1$ dimensional unit sphere.

\begin{figure}[!hbt]
\begin{center}
\includegraphics[width=13cm]{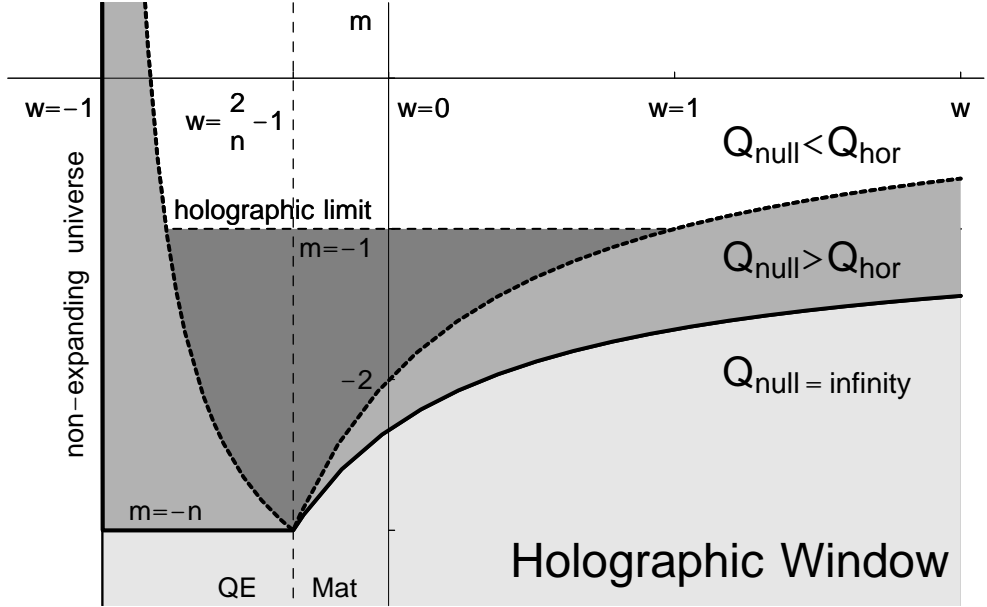}
\end{center}
\caption{{\small The solid line bounds the parameter space region which has a finite value
for $Q_{null}$. The doted line represents the points in the space of parameters with
the same value for both integrals $Q_{null}=Q_{hor}$; we see that regardless of the
space dimension this line passes for the points $(\omega=0,m=-2)$ (dust dominated
universe, density of matter) and ($\omega=1,m=-1$) (extreme matter dominated universe,
density of horizon area). The dashed line is the holographic limit (see text); this
line cuts the adiabatic limit for $\omega=1$.}}
\label{figura1}
\end{figure}

We return to the integral over the null sections (\ref{3});
substituting (\ref{2}) in the integral and defining the
dimensionless variable $x=t/t_{0}$ we obtain
\be\la{7}
Q_{null}=Q_{hor} I(\alpha, n, m),
\ee
where for $\alpha>0$, $I(\alpha, n, m)$ is given by
\be\la{8}
I=\int_{0}^{1} {n \over \alpha} \Big( x^{1-1/\alpha}-x \Big)^{n-1} x^{m} dx=
n B( \alpha(m+n)-n+1,n);
\ee
$B(a,b)$ is the Euler Beta function \ci{abramowitz}
For the case of $\alpha<0$ we have
\be\la{9}
I=\int_{0}^{1} {n \over -\alpha} \Big(x- x^{1-1/\alpha} \Big)^{n-1} x^{m} dx=
n B( -\alpha (m+n), n).
\ee
The integral over the spatial section $Q_{hor}$ is finite if there are not singularities
in the region bounded by the corresponding horizon, what has been implicitly
assumed in (\ref{6}). We cannot say the same for the integral over the past light cone
because it extends until the big bang, a singular point; so the infinite values for
$Q_{null}$ are controlled by the singular behavior of the  Euler Beta function.
In order to avoid this infinities
both arguments of the Beta function must be positive restricting the parameter space
of the cosmological model. One of them is the spatial dimension that obviously satisfy
this requirement. For the other argument we have the restriction
\begin{eqnarray}\la{10}
\alpha>0 &\quad& m>\Big(1 -{1 \over \alpha}\Big) (1-n)-1,\nonumber\\
\alpha<0 &\quad& m>-n,  \,\forall \alpha.
\end{eqnarray}
We note that for the accelerated models the condition $m>-n$ ensures that $Q_{null}$ does
not diverge for $t_{0}=0$, in spite of that for this type of models there is not particle
horizon because all the matter of the universe is causally connected in the big bang.
The region of the parameter space that gives finite values for $Q_{null}$ is represented
in Fig. \ref{figura1}.

The Beta function is symmetric in its arguments $B(a,b)=B(b,a)$;
because the spatial dimension $n$ is an integer, this symmetry
applied to (\ref{7}) will be consistent with a
physical situation, if the other argument, that will play the role of
a spatial dimension in a dual case, is also an integer;
but when both arguments are integers we can relate the quotient between the two
integrals to a combinatorial number,
\begin{eqnarray}\la{11}
\alpha>0 &\quad& {Q_{hor} \over Q_{null} } =
{1 \over n B(\alpha(m+n)-n+1,n)}=
{\alpha(m+n) \choose n}     ,\nonumber\\
\alpha<0 &\quad& {Q_{hor} \over Q_{null} } =
{1 \over n B(-\alpha(m+n),n)}=
{-\alpha(m+n)+n-1 \choose n}.
\end{eqnarray}
The upper entry of the binomial must be higher or equal
than the lower one which translates in
$m\geq -(1-1/\alpha) n$ for $\alpha>0$, and $m\geq -1/\alpha-n$ for $\alpha<0$.
With this conditions the symmetry induced on the quotient of the integrals by the
symmetry of the Beta function is nothing but the property of the binomial coefficients
${a \choose b}={a \choose a-b}$, which implies a duality
between different integrals for different cosmological models.
The two scenarios characterized by $(n,\alpha ,m)$ and $(\bar n,\bar \alpha ,\bar m)$
are dual if $\bar \alpha( \bar m +\bar n)=n+\bar n$
for $\alpha>0$, and $-\bar \alpha( \bar m +\bar n)=n+1$ for $\alpha<0$, and the same
relations for both cases interchanging the barred parameters with the unbarred ones.
This is the case for a strict inequality. When both sides of the previous relations
are equal we have $Q_{null}=Q_{hor}$ which is also represented in the Fig.
\ref{figura1}. For ordinary matter ($\alpha>0$) the line is given by $m=-2/(1+\omega)$
independent of the number of spatial dimensions; this is the exponent for the density
of entropy if we impose an adiabatic evolution. In this case two cosmological
densities will be dual if $m(1+\omega)=-2=\bar m(1+\bar \omega)$ for all $n=\bar n$.
In Table \ref{tabla5} we give some examples of dual flows for different densities in
different cosmological models; we have restricted to universes dominated by
dust, radiation and extreme matter,
$\omega=0,1/n,1$ respectively, and to densities of
volume, horizon area and mass, $m=0,-1,-2$
respectively (see below).

\begin{table}[!htb]
\begin{tabular}{|c|}
\hline
  $ \Bigg\{ \displaystyle { ( \ \omega =\frac{1}{2}, \ m=0, \ n=2 \ )
  \atop ( \bar \omega =0, \ \bar m=-1, \ \bar n=4 \ ) } \Bigg\} $
  $\quad \Rightarrow \quad \displaystyle \frac{Q_{hor}}{Q_{null}}=
  \displaystyle {6 \choose 2}=15 \quad $\\
\hline
\hline
  $ \Bigg\{ \displaystyle { ( \ \omega =\frac{1}{3}, \ m=0, \ n=3 \ )
  \atop ( \bar \omega =0, \ \bar m=-1, \ \bar n=3 \ ) } \Bigg\} $
  $\quad \Rightarrow \quad \displaystyle \frac{Q_{hor}}{Q_{null}}=
  \displaystyle {6 \choose 3}=20 \quad $\\
\hline
\hline
  $ \Bigg\{ \displaystyle { ( \ \omega =0, \ m=0, \ n=3 \ )
  \atop ( \bar \omega =0, \ \bar m=0, \ \bar n=6 \ ) } \Bigg\} $
  $\quad \, \Rightarrow \quad \displaystyle \frac{Q_{hor}}{Q_{null}}=
  \displaystyle {9 \choose 3}=84 \quad $\\
\hline
\hline
  $ \Bigg\{ \displaystyle { ( \ \omega =0, \ m=-2, \ n=\bar n \ )
  \atop ( \omega =1, \ m=-1, \ \bar n \ ) } \Bigg\} $
  $\, \, \Rightarrow \quad \displaystyle \frac{Q_{hor}}{Q_{null}}=
  \displaystyle {n \choose n}=1 \quad \ $\\
\hline

\end{tabular}
\caption{{\small
Three pairs of dual scenarios and a continuous family of them.  }}

\label{tabla5}
\end{table}

Let us apply the general result (\ref{7}) to the three variables that enter in
the first law of thermodynamics namely energy, volume and entropy, its respective
densities scaling with different exponents.
For the energy density the scale law is given by $\rho \propto t^{-2}$
so that $m=-2$ for
all $\omega$ and $n$. In Fig. \ref{figura1} we observe that the line $m=-2$ intersects the
bound (\ref{10}) at $\omega=1-2/n$; that is for higher values of $\omega$
the amount of energy living in the past light cone diverges.
This does not means that an infinite amount of energy
influences the origin in this scenario because
what is observed in the origin is the red shifted
residue of the energy living in the
light sheet. However if we pretend to construct a null foliation of the universe,
the Cauchy data defined on the light sheet are globally infinite
\footnote{Our result can help to study the influence of global
variables \cite{ellis} that give rise to paradoxes like the Olbers'.}.
In general, the relation between the energy
that we see, that is the energy living on our past null cone, and the energy inside
the actual horizon for the decelerated models is given by
\be\la{12}
{ M_{null} \over M_{hor} } = I(\alpha, n, m=-2)=
n B \Big( {n(1- \omega) -2 \over n(1+ \omega) -2 } , n \Big);
\ee
taking $n=3$ to be concrete we obtain
\be\la{13}
{ M_{null} \over M_{hor} } = { \mid(1+3\omega)\mid^{3} \over (1+\omega) (1-3\omega) },
\ee
valid also for the accelerated models. We can see that the
previous relation is symmetric with respect to $\omega=\omega_{c}=-1/3$. The energy
on the light cone diverges both for $\omega=1/3$ and when $\omega \rightarrow -1$ ( $\omega$
strictly $-1$  is a de Sitter universe that we do not study). For $\omega=0$,
a dust model, the same amount of energy resides in both sections;
this is the point $(\omega=0,m=-2)$ of the line $Q_{null}=Q_{hor}$ in the
Fig. \ref{figura1}. For $\omega \in
(-2/3,0)$, $ M_{null} / M_{hor}  <1$, with $ M_{null} / M_{hor} =0$ for
$\omega= -1/3$ due to the infinite size of the horizon.

For the volume we note that its density scales trivially with $m=0$; then, by using
(\ref{7}) the volume observed is finite for any $\omega>-1$. Also by inspection
of the Fig. \ref{figura1} we see that $m=0$ intersects the line $Q_{null}=
Q_{hor}$ so that the volume in the past light cone is higher than the volume
of the region bounded by the horizon if
\be\la{14}
-1 < \omega < {2 \over n(n+1) } -1,
\ee
which is always into the accelerated regime. For the other values of $\omega$
the volume observed is smaller $V_{null}<V_{hor}$.

The entropy can be also analyzed using the general result (\ref{7}). We can reduce
the size of the allowed parameter space by imposing the holographic principle on
the entropy content of the universe. We must be sure that the number of degrees of
freedom living in a volume with $n$ dimensions can be mapped on a hypersurface
of dimension $n-1$. The way in which these bits of information are projected in the
hypersurface is not known in general, but particular cases can be handled succesfully
\ci{maldacena}. Also there are different ways to choose the type of the screens in
which must be projected the relevant degrees of freedom of the bulk.
In cosmology the holographic principle is implemented mainly with the screens
being a sort of  cosmological horizon. There are three different cosmological
horizons to deal with, namely particle \ci{fischler},
event \ci{bekenstein} and apparent horizon \ci{bousso}. The three types of
horizons share the property that its size scales linearly with the cosmic time. According
to the holographic principle the maximum entropy in the corresponding horizon
is bounded by $S_{hor}^{max} = A_{hor}/4 G$ where $G$ is the Newton constant
and $A_{hor}$ is the horizon area that scales with cosmic time as $t^{n-1}$; if
the entropy in the bulk must have enough room on the horizon for all cosmic times
its entropy density cannot scale faster than $t^{-1}$; otherwise there will be
a moment at which the entropy in the bulk will be higher than the allowed amount of it
on the horizon. We have encountered the necessary condition than the
density of cosmological entropy $s(t)=Pt^{m}$ must grow with $m<-1$;
in order to obtain a sufficient condition we can not forget the constant $P$
that is the scale of the corresponding integral.
$m=-1$ gives the holographic limit in the Fig. \ref{figura1}. If the
expansion is adiabatic, the line $S_{null}=S_{hor}$ cuts the holographic limit
at $\omega=1$ reproducing the Fischler-Susskind limit \ci{fischler}. We see that
the value $\omega=1$ that corresponds to the stiffest equation of state for
the fluid is allowed only marginally and saturates the holographic bound as
was obtained in \ci{banks}.

We have not a clear understanding of the basic reason behind
the apparence of the Euler Beta function
in the study of physical density integrals on the  null sections of
flat expanding isotropic universes; we suspect however, that all those
conditions are under the results obtained and that can be related with the conformal
symmetry inherent of this cosmological model. In fact the Euler Beta function
seems to encode the germs of string/M theory \ci{veneziano} and our universe
is a consequence of this ultimate theory.

\section*{Acknowledgements} \nonumber
We thank L.~J.~Boya, M.~Asorey and J.~Sesma for helpful comments.
This work was supported by MCYT (Spain), grant FPA2000-1252.

\end{document}